\documentclass[preprint,showpacs,preprintnumbers,amsmath,amssymb,natbib]{revtex4}

\usepackage{graphicx}
\usepackage{epsfig}		
\usepackage{dcolumn}
\usepackage{bm}
\def\0{\mbox{\tiny $0$}}
\def\1{\mbox{\tiny $1$}}
\def\2{\mbox{\tiny $2$}}
\def\3{\mbox{\tiny $3$}}
\def\4{\mbox{\tiny $4$}}
\def\5{\mbox{\tiny $5$}}
\def\6{\mbox{\tiny $6$}}
\def\7{\mbox{\tiny $7$}}
\def\8{\mbox{\tiny $8$}}
\def\9{\mbox{\tiny $9$}}

\def\R{\mbox{\tiny $R$}}

\def\B{\mbox{\tiny $B$}}

\def\mi{\mbox{\tiny $-$}}

\def\bb#1{\mbox{\footnotesize $(#1)$}}

\begin{document}

\title{Stability of mass varying particle lumps}

\author{A. E. Bernardini}
\affiliation{Departamento de F\'{\i}sica, Universidade Federal de S\~ao Carlos, PO Box 676, 13565-905, S\~ao Carlos, SP, Brasil}
\email{alexeb@ufscar.br}
\author{O. Bertolami}
\affiliation{Instituto Superior T\'ecnico, Departamento de F\'{\i}sica, Av. Rovisco Pais, 1, 1049-001, Lisboa, Portugal}
\email{orfeu@cosmos.ist.utl.pt}
\altaffiliation[Also at]{~Instituto de Plasmas e Fus\~{a}o Nuclear, IST, Lisbon.}

\date{\today}

\begin{abstract}
The theoretical description of compact structures that share some features with mass varying particles allows for a simple analysis of the equilibrium and stability for massive stellar bodies.
We investigate static, spherically symmetric solutions of Einstein equations for a system composed by nonbaryonic matter (neutrinos or dark matter) which forms stable structures through attractive forces mediated by a background scalar-field (dark energy).
Assuming that the dark matter, or massive neutrinos, consist of a gas of weakly interacting particles, the coupling with the scalar field is translated into an effective dependence of the mass of the compounding particle on the radial coordinate of the curved spacetime.
The stability analysis reveals that these static solutions become dynamically unstable for different Buchdahl limits of the ratio between the total mass-energy and the stellar radius, $M/R$.
We also find regular solutions that for an external observer resemble Schwarzschild black-holes.
Our analysis leaves unanswered the question whether such solutions, which are both regular and stable, do exist.
\end{abstract}

\date{\today}
\maketitle

\section{Introduction}

The search for the mechanism responsible for the onset of the accelerating phase of the Universe continues to stimulate interesting and fascinating discussions on many key issues in astrophysics and cosmology \cite{Zla98,Wan99,Ste99,Bar99,Ber00}.
In the context of models account for the accelerated expansion of the Universe, a nontrivial coupling between the nonbaryonic matter sector and the sector responsible for the acceleration of the Universe has often been considered \cite{Ame07,Wet07,Ber07D}.
A possible trigger for the acceleration arises through the interaction of the quintessence field with a matter component whose mass grows with time.
This matter component is sometimes identified with dark matter \cite{Ame02,Cal03,Mot04},
neutrinos \cite{Far04,Bro06A,Bja08,Ber08A,Ber08B,Han09,Pas09}, or is unified with dark energy as in the Chaplygin gas model \cite{Kam02,Bil02,Ber02,Ber03,Ber04,Ber05B}.

Generically speaking, nonbaryonic matter interacting adiabatically with a background scalar field gives origin to the so-called mass varying mechanism, originally conceived to address the dark matter issue, and later reformulated to study cosmological neutrinos \cite{Far04,Han09}.
In its simplest realization, it consists in considering a scalar field effective potential with an implicit dependence on the neutrino mass.
Due to the constraint imposed by the scalar field, the mass eventually grows close to its present value, when neutrinos form a nonrelativistic (NR) fluid and the interaction with the scalar field locks the mass evolution.
The potential energy of the dark energy component becomes then the dominant contribution to the Universe's energy density and the cosmic acceleration ensues.
However, in models with increasing particle mass, the homogeneous cosmological solution is usually unstable to perturbations.
In fact, several forms of coupling of mass varying particles to quintessence fields lead to instabilities associated with an imaginary speed of sound for the fluid in the NR regime.
Such instabilities result in the exponential growth of cosmological perturbations \cite{Bea08}.
The Universe becomes then inhomogeneous and these particle lumps form denser structures.

In this work we quantify the equilibrium and stability conditions for static, spherically symmetric objects that share some features with cosmological scenarios with mass varying particles.
These exotic astrophysical objects are compact lumps of nonbaryonic matter, held together by gravity and the attractive force mediated by the background scalar field.
The static mass varying behavior is achieved from the dependence on the scalar field radial coordinate.
In our analysis, we shall consider that such stellar structures are described by the Schwarzschild's solution of a sphere with constant energy density, $\rho$, and pressure, $p$, which drops from its central value to vanish at the boundary.
We shall determine the equilibrium conditions between the total mass-energy, $M$, and the spherical radius, $R$.
Actually, in our approach the variable mass behaves like an additional extensive thermodynamic degree of freedom, due to its dependence on the scalar field, being therefore like a chemical potential from the thermodynamic point of view.
Under adiabatic conditions, the thermodynamic pressure and its explicit dependence on $r$ are given in terms of $\rho\bb{r}$, and can be explicitly computed from the Tolman-Oppenheimer-Volkoff (TOV) equations for the hydrostatic equilibrium \cite{Tol39,Vol39}.
This approach justifies the assumption of a specific analytical dependence of the variable mass on the radial coordinate of the massive object, $m\bb{r}$, at a given distance from the center.
The stability analysis reveals that our static solutions become dynamically unstable for different Buchdahl limits of the ratio between total mass-energy and stellar radius, $M/R$.
Assuming that such compact structures may describe massive neutrino lumps, we also verify how the structure formation could eventually modify the cosmological predictions for the neutrino masses.
In particular, we find that there are some physically acceptable solutions which give rise to stable lumps.
Our analysis leaves open the issue related with stable regular solutions that closely reproduce the conditions for the formation of Schwarzschild black holes.

This work is organized as follows. In section II, we review well-known thermodynamic properties for neutral matter in order to properly formulate a form of the  first law of thermodynamics that is consistent with the mass varying mechanism.
In section III, we review the fundamental aspects of the mass varying mechanism in the Friedman-Robertson-Walker (FRW) cosmological scenario.
In the context of structure formation, we establish the connection among the energy density $\rho$, the variable mass $m$ and the particle density $n$, for an adiabatic system of mass varying particles in a compact structure.
The corresponding equilibrium conditions and the mass defect are determined in section IV.
In section V, we perform the analysis of stability of the equilibrium configurations and obtain the conditions for neutrino lumping.
We also verify how the cosmological predictions for neutrino masses are modified.
Finally, we summarize our findings and discuss their implications in section VI.

\section{General thermodynamic properties for neutral matter}

In order to analyze the cosmological problem and the theory of stellar evolution, it is necessary to get information on the interactions among the particles which make up the astrophysical bodies.
Regardless the specific nature of these forces, a common property is the additivity of the interaction energy for a macroscopic system: if the system is divided into macroscopic parts, then the interaction energy between these parts will be negligibly small, and for short-range forces we can introduce the concept of specific energy $\epsilon = E/N = \rho/n$, where $\rho$ is the energy density, the total energy, $E$, in a volume, $V$, and $n$ is the density  of particles to which we relate the energy and $N$ is the total number of particles, $N = n V$.

Furthermore, the main manifestation of short-range interactions in a macroscopic system is a nonvanishing pressure $p$.
The pressure is the quantity which allows for defining the interaction force between two parts of a system and depends only on the state of the matter, $p = p(n,\,s)$, i. e. on the density of particles $n$ and on the specific entropy $s = S/N$.
Finally, we have to introduce other two thermodynamic intensive variables: the temperature, $T(n,\,s)$, and the chemical potential, $\mu(n,\,s)$.

The macroscopic version of the first law of thermodynamics leads to the familiar expression which relates the abovementioned quantities
\begin{equation}
\mbox{d}E = - p(V,\,S)\,\mbox{d}V + T(V,\,S)\,\mbox{d}S + \sum_j \mu_j \,\mbox{d}N_j,
\label{star01}
\end{equation}
where the indices $j$ correspond to different types of particles.
Equally fundamental is the conservation of total number of particles.
Suppose a fluid element $V$ whose moving walls are attached to the fluid so that no particles flow in or out.
As the fluid element moves, its volume changes, but the total number of enclosed particles remains fixed,
\begin{equation}
\mbox{d}(\sum_j N_j) = 0,
\label{star02}
\end{equation}
which can be expressed as
\begin{equation}
\mbox{d}N = \mbox{d}(n\, V) = 0,
\label{star03}
\end{equation}
for neutral matter, i. e. for fluid made up by a single type of particle.
In this case, the volume occupied by a particle can be expressed as $V = 1/n$, the corresponding entropy by $S = N\,s = n\,V\,s = s$, and the thermodynamic relation from Eq.~(\ref{star01}) reads,
\begin{equation}
\mbox{d}\epsilon = \mbox{d}\left(\frac{\rho}{n}\right) =  - p(n,\,s)\,\mbox{d}\left(\frac{1}{n}\right) + T(n,\,s)\,\mbox{d}s,
\label{star04}
\end{equation}
which sets
\begin{eqnarray}
\mbox{d}\rho &=& \left(\frac{\partial\rho}{\partial n}\right)_s \mbox{d}n + \left(\frac{\partial\rho}{\partial s}\right)_n \mbox{d}s
= \frac{\rho(n,\,s) + p(n,\,s)}{n} \mbox{d}n + n\,T(n,\,s)\,\mbox{d}s.
\label{star05}
\end{eqnarray}
The analysis further simplifies the fluid, already assumed to have the same composition, has the same entropy per particle, $s$, and is in a state of adiabatic flow,
so that pressure and energy density are related by
\begin{equation}
n \left(\frac{\partial\rho}{\partial n}\right)_s  = \rho(n,\,s) + p(n,\,s).
\label{star06}
\end{equation}

The influence of the interactions on the thermodynamics and the assumptions about the spectrum of rest masses in the context of neutral matter stellar structures have been recurrently examined the literature.
In the following analysis, we shall consider neutral matter with no color, electroweak and electric charges such as (cold) dark matter and NR neutrinos.
These considerations will allow us to describe interactions of neutral particles within stellar objects whose evolution is driven by the coupling to the dark energy scalar field.

\section{The mass varying mechanism}

In the so-called mass varying mechanism \cite{Far04,Pec05,Bro06A,Bja08} a coupling between a relic particle and a light scalar field is introduced.
This fluid is identified with dark energy or dark energy plus dark matter \cite{Ber08A,Ber08B}.
As a consequence of this new interaction, the mass of the relic particle is generated from the vacuum expectation value of the scalar field and becomes linked to its dynamics, that is $m \equiv m\bb{\phi}$ for such a relic particle.

In the usual MaVaN framework, neutrinos are coupled to a light scalar field which is identified with the dark sector.
Presumably, the neutrino mass $m_{\nu}$ has its origin on the vacuum expectation value (VEV) of the scalar field and, naturally, its behavior is governed by the dependence of the scalar field on the scale factor.
Given a particle statistical distribution $f\bb{q}$, where $q \equiv \frac{|\mbox{\boldmath$p$}|}{T_{\0}}$, $T_{\0}$ being the neutrino background temperature at present, in the flat FRW cosmological scenario, the corresponding energy density and pressure can be expressed by
\begin{eqnarray}
\rho\bb{a, \phi} &=&\frac{T^{\4}_{\nu \0}}{\pi^{\2}\,a^{\4}}
\int_{_{0}}^{^{\infty}}{\hspace{-0.3cm}dq\,q^{\2}\, \left(q^{\2}+\frac{m^{\2}\bb{\phi}\,a^{\2}}{T^{\2}_{\0}}\right)^{\1/\2}\hspace{-0.1cm}f\bb{q}},\\
p\bb{a, \phi} &=&\frac{T^{\4}_{\0}}{3\pi^{\2}\,a^{\4}}\int_{_{0}}^{^{\infty}}{\hspace{-0.3cm}dq\,q^{\4}\, \left(q^{\2}+\frac{m^{\2}\bb{\phi}\,a^{\2}}{T^{\2}_{\nu \0}}\right)^{\mi\1/\2}\hspace{-0.1cm} f\bb{q}},~~~~ \nonumber
\label{gcg01}
\end{eqnarray}
where sub-index $0$ denotes present-day values, such that $a_{\0} = 1$.
It can be easily shown that
\begin{equation}
m\bb{\phi} \frac{\partial \rho\bb{a, \phi}}{\partial m\bb{\phi}} = (\rho\bb{a, \phi} - 3 p\bb{a, \phi}).
\label{gcg02}
\end{equation}
From the dependence of $\rho$ on $a$, one can obtain the energy-momentum conservation for the mass varying fluid,
\begin{equation}
\dot{\rho}\bb{a, \phi} + 3 H (\rho\bb{a, \phi} + p\bb{a, \phi}) =
\dot{\phi}\frac{d m\bb{\phi}}{d \phi} \frac{\partial \rho\bb{a, \phi}}{\partial m\bb{\phi}},
\label{gcg03}
\end{equation}
where $H = \dot{a}/{a}$ is the expansion rate of the Universe and the {\em dot} denotes differentiation with respect to cosmic time ($^{\cdot}\, \equiv\, d/dt$).
It is important to emphasize that the coupling between relic particles and the scalar field as described by Eq.~(\ref{gcg02}) is restricted to times when they are NR, i. e. $\frac{\partial \rho\bb{a, \phi}}{\partial m\bb{\phi}} \simeq n\bb{a} \propto{a^{\mi\3}}$ \cite{Far04,Bja08,Pec05}.
In opposition, as long as particles are relativistic ($T\bb{a} = T_{\nu \0}/a >> m\bb{\phi\bb{a}}$), the decoupled fluids should evolve adiabatically since the strength of the coupling is suppressed by the relativistic increase of the pressure ($\rho\sim 3 p$).
The mass varying mechanism is essentially driven by Eq.~(\ref{gcg02}), which translates the dependence of $m$ on $\phi$ into a dynamical behavior.

However, the mass varying mechanism can play a salient role in the structure formation.
It can be assumed that cosmological neutrinos, dark matter, or some generic form of nonbaryonic matter, compose a gas of weakly interacting particles.
The mass of these particles depends on the value of a slowly varying classical scalar field.
This dynamics is similar to that of the so-called {\em cosmon} fields which can lead to the formation of {\em cosmon} lumps \cite{Wet94,Bea08}.

{\em Cosmon} lumps are bound objects, for which the scalar and gravitational fields combine to form nonlinear solutions of their field equations.
Such solutions have been discussed in detail in Refs. \cite{Wet02,Wet08,Tet08} for situations where the influence of the {\em cosmon} potential is negligible.
Their characteristic features are: (i) lumps are designated by their mass and a suitably defined {\em cosmon} charge; (ii) the black-hole event horizon disappears for all solutions with nonzero {\em cosmon} charge; (iii) for small {\em cosmon} charge, the solutions closely resemble black-hole solutions when seen from the outside, and the horizon is replaced by a sharp transition to a region of high red-shift.

For classical particles, the action of the {\em cosmon} coupled to mass varying particles ($\nu$ or DM) can be written as \cite{Wet08}
\begin{equation}
S = \int{\mbox{d}^{\4} x \sqrt{-g} \left[\frac{R}{16 \pi G} + \frac{1}{2}g^{\mu\nu}\partial_{\mu}\phi \partial_{\nu}\phi + V\bb{\phi} + \rho\bb{\phi}\right]}
\label{gcg03B}
\end{equation}
where $G$ is the Newton constant, $R$ is the Ricci curvature and $\rho\bb{\phi}$ is the corresponding mass varying particle density.
For static, spherically symmetric solutions, one employs the Schwarzschild metric given by $\mbox{d}s^{\2} = -B\bb{r} \mbox{d}t^{\2} + A\bb{r} \mbox{d}r^{\2} + r^{\2}\left(\mbox{d}\theta^{\2} + \sin^{\2}(\theta)\mbox{d}\varphi^{\2}\right)$, where $A = B^{-\1} = (1 - 2 G M /r)^{-\1}$, for which we set $G=1$, and $r$ is the radial coordinate.
From the field equations we can write
\begin{eqnarray}
\frac{\mbox{d}^{2}\phi}{\mbox{d} r^{2}} + \left(\frac{2}{r} + \frac{B^{\prime}}{2 B} - \frac{A^{\prime}}{2 A}\right)
\frac{\mbox{d}\phi}{\mbox{d} r}
&=&
A \left (\frac{\mbox{d} V\bb{\phi}}{\mbox{d} \phi}  + \frac{\partial \rho\bb{\phi}}{\partial \phi} \right)\nonumber\\
&=&
A \left (\frac{\mbox{d} V\bb{\phi}}{d \phi}  + \frac{\mbox{d} \ln{m\bb{\phi}}}{d \phi} (\rho - 3 p)\right),
\label{gcg03C}
\end{eqnarray}
from which the explicit dependence of $\phi$ on $r$ is obtained.

Assuming the adiabatic approximation at cosmological scales \cite{Bea08}, the cosmological stationary condition \cite{Far04,Bro06A,Bja08,Ber08B} can be applied to Eq.~(\ref{gcg03C}) in order to suppress its right hand side.
The analytical dependence of the scalar field on the radial coordinate effectively disappears for large values.
Faraway from the concentration of matter, i. e. for $r$ larger than the compact structure radius, one recovers the background of a flat FRW Universe.
In this case, considering back the usual cosmological time-dependent terms, the Eq.~(\ref{gcg03C}) should reduce to the energy conservation equation,
\begin{equation}
\ddot{\phi} + 3 H \dot{\phi} + \left (\frac{\mbox{d} V\bb{\phi}}{d \phi}  + \frac{\mbox{d} \ln{m\bb{\phi}}}{d \phi} (\rho - 3 p)\right)
= 0,
\label{gcg08}
\end{equation}
and the stationary condition could be equivalently assumed for the static approach, i. e.
\begin{equation}
\frac{\mbox{d} V\bb{\phi}}{d \phi}  + \frac{\mbox{d} \ln{m\bb{\phi}}}{d \phi} (\rho - 3 p)
= 0.
\label{gcg08C}
\end{equation}
This equation does not depend on the relation of $\phi$ with spacetime coordinates or metric components, that is, the radial coordinate, $r$, of a curved region or the scale factor, $a$.
For this reason, the simplest scenario is obtained from the extension of the cosmological stationary condition to a static configuration in Eq.~(\ref{gcg03C}).
The adiabatic regime depicted from the gradual change expressed by Eq.~(\ref{gcg08C}) can be compared to the minimum of the effective potential of a chameleon models \cite{Kho04,Kho03}, where the chameleon scalar field are modified through spatial variations in the interior of massive bodies.
In addition, the corresponding energy momentum conservation law is exactly what one encounters in the Einstein frame of an scalar-tensor gravity model \cite{Ame00,Car04}.
In Ref. \cite{Bea08}, this adiabatic regime and its domain of validity is reviewed in a general context.

Introducing the abovementioned Schwarzschild expressions for $A\bb{r}$ and $B\bb{r}$, and assuming the stationary condition, Eq.~(\ref{gcg03C}) can be simplified as
\begin{eqnarray}
\frac{\mbox{d}^{2}\phi}{\mbox{d} r^{2}} + \left[\frac{2 - 8(M/R)(r^{\2}/R^{\2})}{r - 2(M/R) (r^{\3}/R^{\2})}\right]
\frac{\mbox{d}\phi}{\mbox{d} r}
&=& 0 ~~~~ (r < R),\nonumber\\
\frac{\mbox{d}^{2}\phi}{\mbox{d} r^{2}} + \left(\frac{2(r - M)}{r - 2 M}\right)
\frac{\mbox{d}\phi}{\mbox{d} r}
&=& 0 ~~~~ (r > R),
\label{gcg03CDM}
\end{eqnarray}
which yields the solution
\begin{eqnarray}
\phi\bb{r} &=& \phi^{in}_{\0} + \frac{\phi^{in}_{\1}}{2M} \ln{\left(1 - \frac{2M}{r}\right)} ~~~~ (r < R),\nonumber\\
\phi\bb{r} &=& \phi^{out}_{\0} + \phi^{out}_{\1}\sqrt{(2M/R^{\3})}
\left[\frac{1}{\sqrt{(2 M /R^{\3})} r} - arc\tanh{(\sqrt{(2M/R^{\3})} r)}\right] ~~~~ (r > R),
\label{gcg03CDMA}
\end{eqnarray}
where $\phi^{in,out}_{\0,\1}$ are constants to be adjusted in order to match the boundary conditions.
In the Newtonian limit where $A\bb{r}\approx B\bb{r} \approx 1$, the constants match one each other so that $\phi^{in}_{\0,\1} \equiv \phi^{out}_{\0,\1} \equiv\phi_{\0,\1}$ and the above solutions reduce to
\begin{equation}
\phi\bb{r} = \phi_{\0} + \frac{\phi_{\1}}{r}.
\label{gcg08D}
\end{equation}

\begin{figure}
\vspace{-1.0 cm}
\centerline{\psfig{file=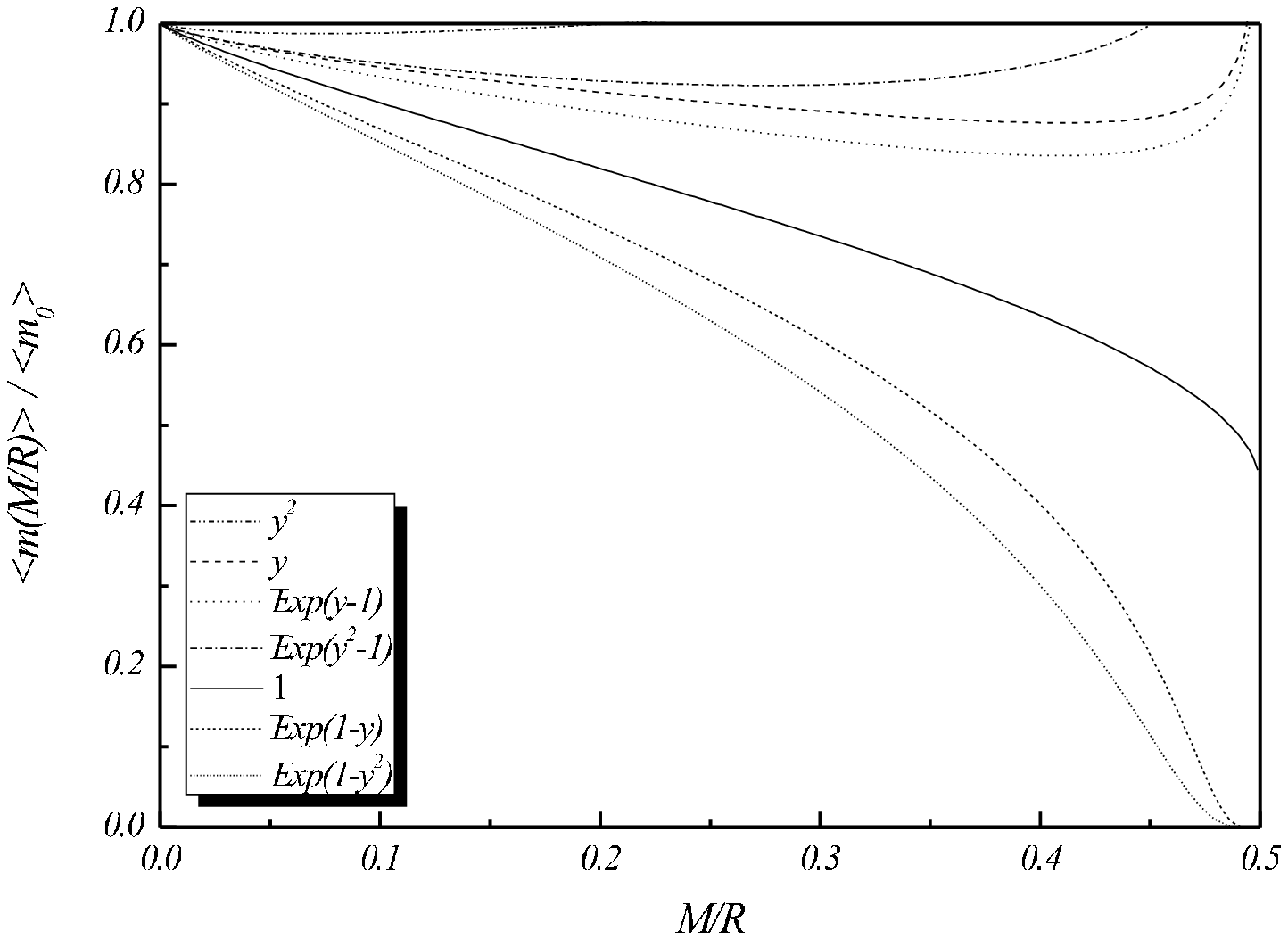, width=14cm}}
\vspace{-1.0 cm}
\caption{Evolution for the variable mass dependence on the radius $r$.
Its dependence is supposed to be dictated by a spherically symmetric distribution of matter due to the coupling with a background scalar field $\phi\bb{r}$.}
\label{Fstar01}
\end{figure}

Given that $\phi$ depends on the radial coordinate for generic curved spaces, we quantify the changes due to our approach through some mass dependencies on $r$ as illustrated in Fig.~\ref{Fstar01}.
Clearly, the form of $m\bb{\phi}$ and of the relation between pressure, $p\bb{r}$, and density, $\rho\bb{r}$, may lead to quite different scenarios.

In the pictorial representation of our results we denote $y = r/R$, where $R$ is the radius of the spherical lump.
Any prescription for $m\bb{r}$ is model dependent, i. e. arbitrary functions for $m\bb{r}$ are equivalent to arbitrary functions for $\phi\bb{r}$ and $m\bb{\phi}$.
In particular, the stationary condition approximation justifies the analytical dependence given by Eq.~(\ref{gcg31}).
However, other approximations could be adopted.

Let us now turn to scenarios involving neutrinos.
Despite its impressive phenomenological success, it is widely believed that the Standard Model (SM) of particle physics is actually only a low-energy approximation of an underlying more fundamental theory.
In this respect, the interplay with the cosmology can be an important guideline to obtain insights on the nature of the more fundamental theory.
Turning to the subject of neutrinos in the SM, the most natural way to explain the smallness of their masses is through the seesaw mechanism, according to which, the tiny masses, $m_{\nu}$, of the usual left-handed neutrinos are obtained via a very massive, $M$, {\em sterile} right-handed neutrino.
The Lagrangian density that describes the simplest version of the seesaw mechanism through the Yukawa coupling between a light scalar field and a single neutrino flavour is given by
\begin{equation}
{\cal L} = m_{LR} \bar{\nu}_L \nu_R + M\bb{\phi} \bar{\nu}_R \nu_R + h.c.,
\label{gcg30}
\end{equation}
where it is shown that at scales well below the right-handed neutrino mass, the following effective Lagrangian density emerges \cite{seesaw}
\begin{equation}
{\cal L} = \frac{m_{LR}^{\2}}{M\bb{\phi}} \bar{\nu}_L \nu_R + h.c..
\label{gcg31}
\end{equation}
Phenomenological consistency with the SM implies that logarithm corrections to the above terms are small. Furthermore from solar, atmospheric, reactor and accelerator neutrino oscillation experiments it is gathered that neutrino masses given by $m_{\nu} \sim m_{LR}^{\2}/M\bb{\phi}$ lie in the sub-$eV$ range.
It is also clear that promoting the scalar field $\phi$ into a dynamical quantity leads to a mechanism in the context of which neutrino masses are time-dependent.
Associating the scalar field to the dark energy field allows linking NR neutrino energy densities to late cosmological times \cite{Far04,Bro06A,Bja08,Ber08A,Ber08B}.
This scenario was implemented through the adiabatic (stationary) approach via Eqs.~(\ref{gcg03C}), (\ref{gcg08}) and (\ref{gcg08C}).
It is evident that this approach is fairly general, and that the same analysis can be performed in the framework of mass varying particles coupled with {\em cosmon} fields in curved spacetime.
Analogously, an approximate solution for a spherically symmetric compact object of radius $R$ with homogeneous
density were already obtained in the quoted chameleon framework \cite{Kho04}

\subsection{Thermodynamics of "noninteracting" mass varying particles}

To describe the connection among the extensive quantities $\rho$, $m$ and $n$ in an adiabatic system of mass varying particles in a star (or any stellar object), the relevant thermodynamic equations are summarized by Eqs.~(\ref{star06})-(\ref{gcg02}).
In fact, in order to get a closed system of equations, an equation of state that relates the pressure in terms of the energy density and the specific entropy, $p = p\bb{\rho, s}$.
Moreover, due to the adiabatic conditions set up, the pressure and its explicit dependence on the radial coordinate, $r$, is simply given in terms of $\rho\bb{r}$, which can be explicitly obtained from the TOV equations for the hydrostatic equilibrium.

By assuming that the hydrostatic pressure, $p = p\bb{\rho\bb{r}}$, leads to a univoquous dependence of $\rho$ on the radial coordinate for a symmetrically spherical distribution of matter, we have to obtain a density $\rho\bb{m\bb{r}, n\bb{r}, r}$ which satisfies the system of partial differential equations given by
\begin{equation}
\frac{\mbox{d}\rho}{\mbox{d}r} = \frac{\partial \rho}{\partial m}\frac{\mbox{d} m}{\mbox{d}r} + \frac{\partial \rho}{\partial n}\frac{\mbox{d} n}{\mbox{d}r} + \frac{\partial\rho}{\partial r},
\label{star13}
\end{equation}
and Eqs.~(\ref{star06})-(\ref{gcg02}).
Eliminating $p$ from Eqs.~(\ref{star06})-(\ref{gcg02}), we obtain
\begin{equation}
4 \rho = m \frac{\partial \rho}{\partial m} + 3 n \frac{\partial \rho}{\partial n},
\label{star14}
\end{equation}
and consequently,
\begin{equation}
\rho\bb{m, n} = \kappa m^{\1-\3 \alpha} n^{\1 + \alpha}
\label{star15}
\end{equation}
is solution of Eq.~(\ref{star13}) once setting the dependence $\alpha \rightarrow \alpha\bb{r}$.
Using once again Eqs.~(\ref{star06})-(\ref{gcg02}), we obtain $\alpha\bb{r} = p\bb{r}/\rho\bb{r}$ and hence the most general solution for the problem is given by
\begin{equation}
\rho = \rho\bb{m\bb{r}, n\bb{r}, r} = m^{\1-\3 \frac{p}{\rho}} n^{\1 + \frac{p}{\rho}},
\label{star16}
\end{equation}
where, for simplicity, we have omitted the explicit dependence of $m$, $n$, $p$ and $\rho$ on $r$, and we have adjusted the arbitrary constant $\kappa$ in order to have $\rho = m \, n$ in the nonrelativistic limit, and $\rho = n^{\4/\3}$ in the ultra-relativistic limit.

For our purpose, a simple and realistic model for a symmetrically spherical stellar object arises from the assumption that the fluid is incompressible: the density $\rho_{\0}$ is a constant out to the surface of the star, $r = R$, after which it vanishes.
Specifying $\rho\bb{r}$ is equivalent to employ the equation of state, since $p\bb{r}$ can be determined from the hydrostatic equilibrium, i. e. from the TOV equations \cite{Tol39},
\begin{equation}
\frac{\mbox{d} p\bb{r}}{\mbox{d}r} = -(\rho\bb{r} + p\bb{r}) \frac{(M \bb{r} + 4 \pi\,r^{\3}\,p\bb{r})}{r (r - 2 M)},
\label{star16A}
\end{equation}
and
\begin{equation}
\frac{\mbox{d} M}{\mbox{d}r} = 4 \pi \, r^{\2} \,\rho\bb{r},
\label{star16B}
\end{equation}
for which we have set Newton's constant, $G = 1$, and $M$ is the total mass of the stellar object.

\begin{figure}
\vspace{-1.0 cm}
\centerline{\psfig{file=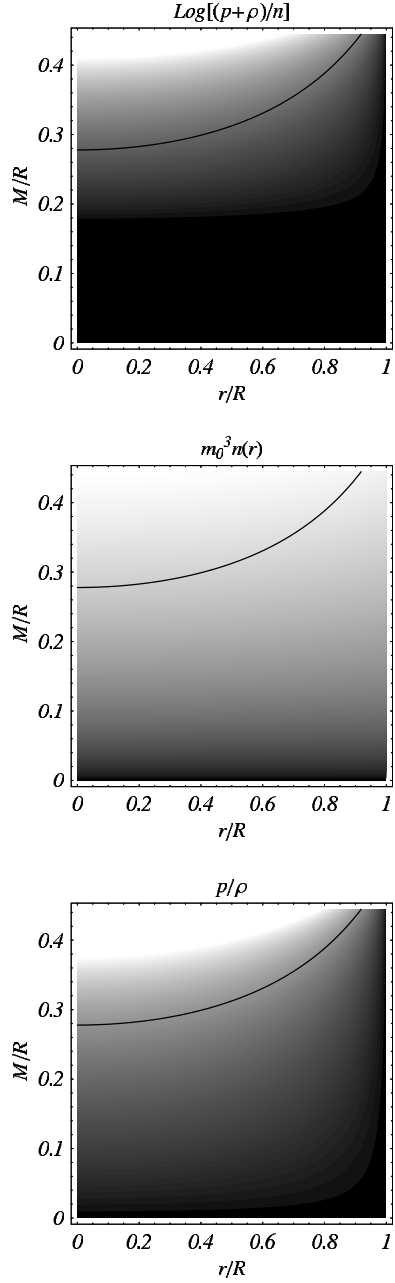, width=18cm}}
\vspace{-1.0 cm}
\caption{Thermodynamic variables for a system composed by mass varying particles, where the mass is assumed to be an extensive thermodynamic degree of freedom.
The plots depict the case for which the mass of the particle is given by its stationary observable value $m\bb{R} = m_{\0}$ on the surface boundary.
Notice that increasing {\em gray level} corresponds to increasing values of the thermodynamic variables, for which the boundary values have been marked for $M/R = 5/18$, the ``soft'' Buchdahl's limit for which $\rho = 3 p$ at $R = 0$.}
\label{Fstar02}
\end{figure}

\begin{figure}
\vspace{-1.0 cm}
\centerline{\psfig{file=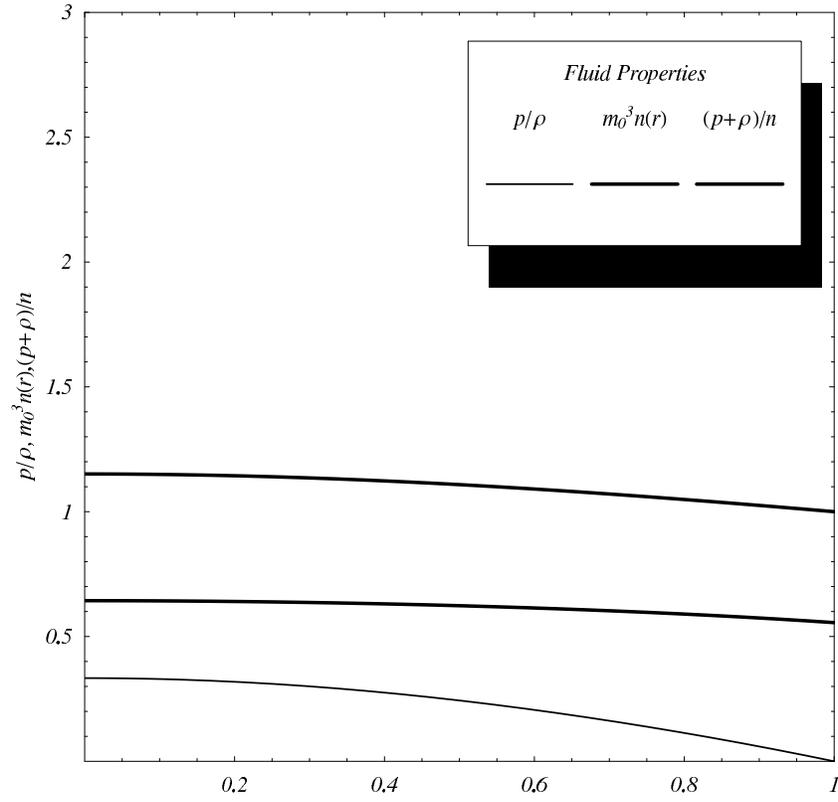, width=11cm}}
\vspace{-1.0 cm}
\caption{The ratio between pressure and energy density $p/\rho$, the density of particles $n$ (dimensionally normalized by $m_{\0}^{\3}$) and the ratio $(p +\rho)/n$.
We assume the ``soft'' Buchdahl's limit for which $\rho = 3 p$ at $R = 0$.
This corresponds to sections of graphs marked on Fig.~\ref{Fstar02}.}
\label{Fstar03}
\end{figure}

Integrating Eq.~(\ref{star16A}) yields
\begin{equation}
\frac{p\bb{r}}{\rho\bb{r}} = \frac{\left[1 - 2(M/R)(r^{\2}/R^{\2})\right]^{\1/\2} - \left[1 - 2(M/R)\right]^{\1/\2}}{ 3\left[1 - 2(M/R)\right]^{\1/\2} - \left[1 - 2(M/R)(r^{\2}/R^{\2})\right]^{\1/\2}}.
\label{star17}
\end{equation}
Notice that the pressure increases near the core of the star, as expected.
Figs.~\ref{Fstar02}-\ref{Fstar03} illustrate the properties of the relevant thermodynamic quantities for the case where their behavior are constrained by Eq.~(\ref{star16}) in the limiting case where the mass of the particle is given by its observable value $m\bb{R} = m_{\0}$.
We compute the ratio between pressure and energy density $p/\rho$, the density of particles $n$ (dimensionally normalized by $m_{\0}^{\3}$), and the ratio $(p +\rho)/n$ that characterizes a {\em pseudo} chemical potential.

Indeed, for a star of fixed radius $R$, the central pressure tends to infinity if the mass exceeds $4R/9$, and exceeds the radiation pressure $(p = \rho/3)$ if the mass exceeds $5R/18$ (solid line in the Fig.~\ref{Fstar02}).

If we try to squeeze a mass greater than the Buchdahl's limit $4R/9$ within a radius $R$, general relativity admits no static solutions \cite{Buc59}; a star shrinks to such size till it eventually turns into a black hole.
This scenario can be modified for astrophysical bodies composed by particles subject to the mass varying mechanism.
Depending on the analytical mass dependence on the cosmological background scalar field modified by the spacetime curvature, the stability of equilibrium configurations can be drastically modified, as we shall discuss in the following section.

\section{Equilibrium and mass defect}

Among various types of equilibrium in stellar and cosmological scenarios, the simplest one is that in which all processes of reciprocal transformation (creation/annihilation) take place at a time scale much faster than the one associated to the flux of external particles and the rate of change of the relevant thermodynamic quantities.
Such a simplified version of thermodynamic equilibrium is consistent with the stellar structure of our approach.
The conditions for lumping of such an exotic nonbaryonic matter plus scalar field fluid is determined by the total binding energy of the system.

The form of the Eq.~(\ref{star16B}) suggests that we interpret $M$ as the total (mass) energy inside a radius $R$, including the rest mass-energy $M_{\0}$, the internal energy of motion $W$ and the (negative) potential energy of self-gravitation $U$.
In the relativistic domain, we have
\begin{equation}
M = M_{\0} + M_{\B} = 4 \pi \int_{\0}^{\R}{r^{\2} \rho\bb{r}\,\mbox{d}r},
\label{star18A}
\end{equation}
where
\begin{equation}
M_{\0} = 4 \pi \int_{\0}^{\R}{r^{\2} m\bb{r} n\bb{r} \sqrt{A\bb{r}}\,\mbox{d}r}.
\label{star18B}
\end{equation}
Following Eq.~(\ref{gcg02}), $M$ from Eq.~(\ref{star18A}) is reduced to $M_{\0}$ (no binding energy and no motion) if $p = 0$.
The difference between the total rest mass-energy and the total energy corresponds to the (positive) binding energy $M_{\B}$ which keeps the stellar structure stable,
\begin{equation}
M_{\B} =  M_{\0} - M = 4 \pi \int_{\0}^{\R}r^{\2}{\left[ \rho\bb{r} - m\bb{r} n\bb{r}\sqrt{A\bb{r}}\right]\mbox{d}r}.
\label{star18C}
\end{equation}

Although it is useful to define $M$ as the total energy, the rest mass-energy $M_{\0}$ is a fundamental quantity in the stability analysis.
In opposition, the internal energy of motion, $W$, and the (negative) potential energy of self-gravitation, $U$, are not particularly relevant, unless in the Newtonian approximation where the relation $M_{\B} \approx U + W$ holds.
The binding energy is sometimes referred to as {\em mass defect} and it corresponds to the energy which is released during the formation of a star from an initially rarefied matter: a typical mechanism for compact structure formation.
When particles are combined into a bound system, an energy which equals the mass defect is emitted in the form of photons, relativistic neutrinos, or gravitational waves.
For this reason, it follows from the physics of this process that the condition for a stable static star originating from diffuse matter distribution is expressed as $M_{\B} > 0$.

\begin{figure}
\vspace{-1.0 cm}
\centerline{\psfig{file=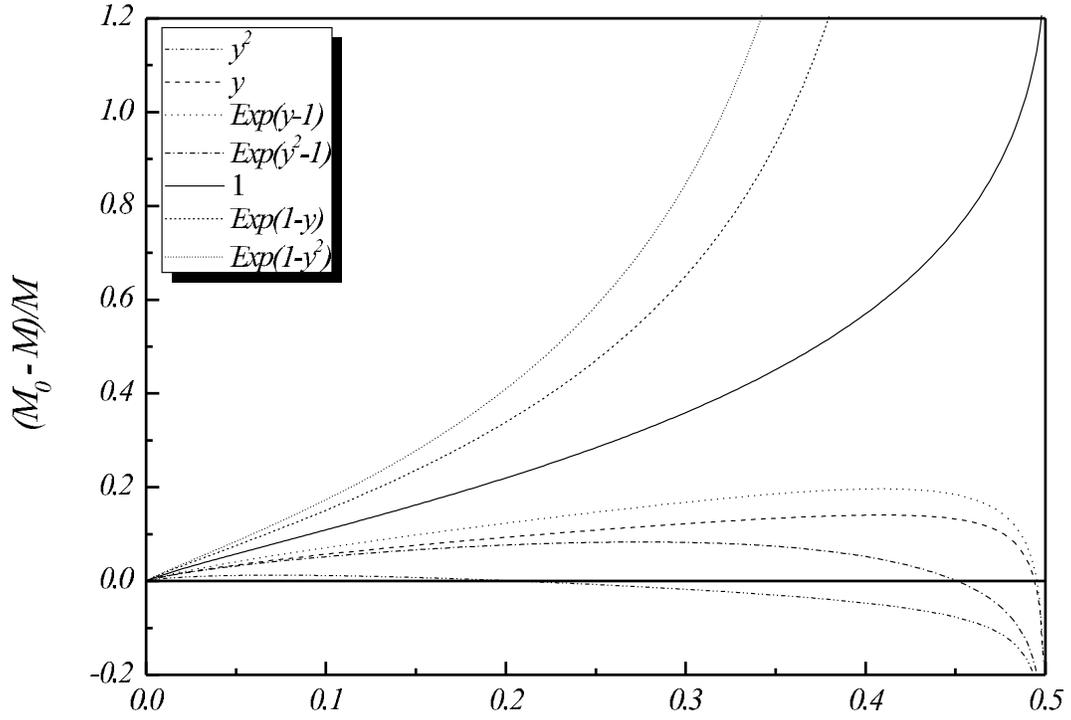, width=14cm}}
\vspace{-1.0 cm}
\caption{Mass defect (binding energy) for several mass varying particle functions for compact objects with uniform energy density.
Given that $\phi$ depends on the radial coordinate $r$ for generic classes of curved spaces, we quantify the modifications due to our approach by some mass dependencies on $r$ as illustrated in the box, where $\mbox{y} = r/R$.}
\label{Fstar04}
\end{figure}

In the Fig.~\ref{Fstar04} we show how the mass varying mechanism does modify the equilibrium condition based on the mass defect criterium for relativistic stellar objects.

Two interesting effects can be described through the mass varying mechanism depending on whether the particle mass increases or decreases inwards towards the center.
From the behavior of the mass defect for compact objects composed by particles with their mass exponentially increasing inwards to the center, we notice that stable configurations are favored relatively to the other analytical dependencies of $m\bb{r}$.
The resulting compact structure has binding energy that increases to infinite as the ratio $M/R$ approaches to its Buchdahl's limit.
When these objects shrink to such a size they inevitably keep on shrinking and eventually give origin to a black hole.

In the absence of the mass varying mechanism, the rest mass-energy $M_{\0}$ is written as $M_{\0} = m \, N$, where $N$ is the total number of particles within the radius $R$, that is
\begin{equation}
N = 4 \pi \int_{\0}^{\R}{r^{\2} n\bb{r} \sqrt{A\bb{r}}\,\mbox{d}r}.
\label{star19}
\end{equation}
The stability curve for this case is described by the solid line depicted in the Fig.~\ref{Fstar04}, for which the binding energy has an upper limit, $M/R < 4/9$, whether one assumes that the pressure at the center cannot go to infinity.
It is important to realize that, in case of mass varying particles, the analysis of the stability conditions in terms of $N$ has to be rephrased in terms of $M_{\0}$.

In this context, decreasing the mass inwards towards the center can also give origin to stable structures up to certain limiting values for $M/R$.
In opposition to the previous result, the equilibrium conditions are achieved for smaller values of  $M/R$, which restricts the existence of neutrino lumps with arbitrarily small masses.
The point here is that the coupling with the background scalar field is crucial in determining the stability conditions.
In this situation, the role of the scalar field potential on the mass varying mechanism requires specific considerations, which not only depend on the cosmological nature of the scalar field (quintessence, cosmon, phanton, Chaplygin gas, etc), but also on the mass generation mechanism.

\section{Stability of equilibrium configurations}

The condition of hydrostatic equilibrium is equivalent to the condition of an extremum of the total energy of a spherically symmetric stellar object for a given number of conserved particles and a given specific entropy.
In fact, the relativistic theory of radial perturbations for nonrotating equilibrium configurations is well understood.
There are two approaches to radial stability: a dynamical one, based on the equations of motion and energy properties of radial perturbations; and a static one, based on mass-radius curves for certain sequences of equilibrium configurations.
From the dynamical approach we can find different ways for studying the normal radial modes of a relativistic stellar model.
From the static approach we obtain the minimum energy corresponding to stable equilibrium, and a maximum to unstable equilibrium.
In the static approach the study of stability does not require additional computations.
It clearly reveals that general relativity catalyzes radial instability in stellar models.

The analysis of stability conditions from the static approach is similar to one when treating mass varying particle systems with thermodynamics governed by Eq.~(\ref{star16}).
However the analysis has to be performed in terms of $M_{\0}$ and not in terms of $N$.

First, we consider how the mass of an equilibrium configuration changes with the addition of particles, brought from infinity, where the energy is $M$.
In other words, we derive the analytical correspondence between $M$ and $M_{\0}$ and, in particular, $\mbox{d}M/\mbox{d}M_{\0}$.
The change in $M$ should not depend on the position in the equilibrium configuration at which the particle is added up.

It can also be shown that for stable equilibrium configurations, $\mbox{d}M/\mbox{d}M_{\0} < 1$ \cite{ZelXX,ThoXX}.
This does not necessarily imply that for equilibrium $M < M_{\0}$ for equilibrium, which is an independent condition.
In fact, the equilibrium solutions with negative binding energy $M_{\B}$ are unstable.
Moreover, the stability against small perturbations is lost at the point of greatest (positive) binding energy, so that the configurations of negative binding energy must lie deep into the unstable region.
It is important to realize that the NR analysis, which is essentially qualitative, allows for solutions of negative binding energy \cite{ZelXX}.
The stability condition is reduced to $\mbox{d}M/\mbox{d}N < m$ for the case where the particle masses does not depend on the radial coordinate, $m\bb{r} = m = m_{\0}$.
For this reason, the usual general relativity analysis of stability is performed in terms of $M$ instead of $N$.
In the case of mass varying particle systems it gives origin to significant deviations from the equilibrium and stability scenarios described in terms of the condition $M < M_{\0}$.

\subsection{Stability of mass varying particle lumps}

\begin{figure}
\vspace{-1.0 cm}
\centerline{\psfig{file=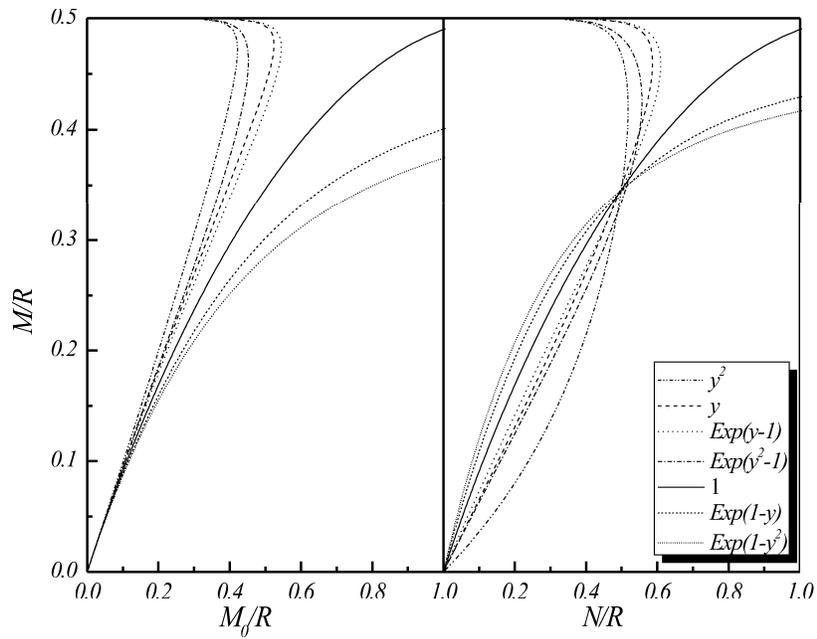, width=14cm}}
\vspace{-1.0 cm}
\caption{Stability curves for compact objects in correspondence with the mass varying mechanism prescription.
We plot $M/R \times M_{\0}/R$ - the right analysis element - and $M/R \times N/R$ - the wrong analysis element - for the case where $m$ is given in units of $m_{\0}$.
They are all in correspondence with the graphs of Fig.~\ref{Fstar06}.}
\label{Fstar05}
\end{figure}

By considering the above discussed static approach for the study of stability, let us describe the stability curves for the considered mass varying particle analytical relationships, and compare them with the analysis in terms of the particle number $N$.
In Fig.~\ref{Fstar05}  we plot the stability curves, i. e. the relationship between the total mass energy, $M$, and the rest mass-energy, $M_{\0}$, and its correspondence with respect the analysis in terms of the total number of particles contained within $R$.

\begin{figure}
\vspace{-1.0 cm}
\centerline{\psfig{file=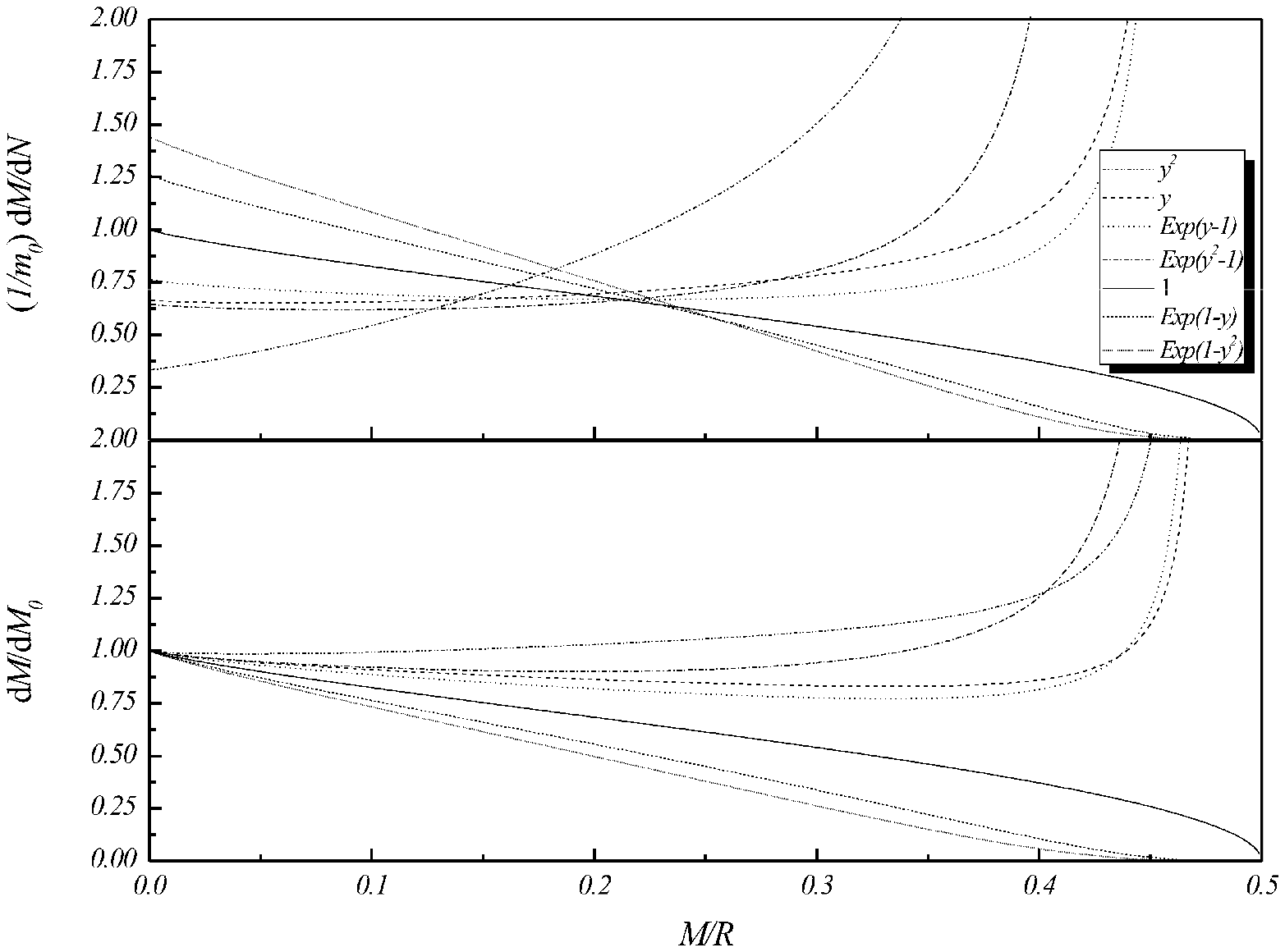, width=14cm}}
\vspace{-1.0 cm}
\caption{Stability condition for compact objects in correspondence with the mass varying mechanism prescription.
We plot $\mbox{d}M/\mbox{d}M_{\0}$ - the right analysis element - and $\mbox{d}M/\mbox{d}N$ - the wrong analysis element - for the case where $m$ is given in units of $m_{\0}$.
Curves are all in correspondence with the graphs of Fig.~\ref{Fstar05}.}
\label{Fstar06}
\end{figure}

In the Fig.~\ref{Fstar06} we plot the stability conditions, i. e. $\mbox{d}M/\mbox{d}M_{\0}$ and $\mbox{d}M/\mbox{d}N$ for the case where $m$ is given in units of $m_{\0}$.

The stable equilibrium configurations are achieved if $\mbox{d}M/\mbox{d}M_{\0} < 1$, and not if
$\mbox{d}M/\mbox{d}N < 1$.

\subsection{Neutrino Lumps}

Neutrinos coupled to the dark energy, an ultra light scalar field lead, in its simplest formulation, to a number of significant phenomenological implications.
Under general conditions, they materialize in a rather model independent way \cite{Far04,Pas09}.
The neutrino mass $m_{\nu}$ is presumably generated from the vacuum expectation value (VEV) of the scalar field $\phi$ whose dynamics is governed by the Universe scale factor $a$, $\phi \equiv \phi(a)$, which turns the neutrino mass into a dynamical quantity.

In most of the scenarios, the neutrinos remain essentially massless until recent times.
Their mass eventually increases close to its present value and their interaction with the background scalar field almost ceases \cite{Ber08A,Ber08B}.
The energy of the scalar field becomes the dominant contribution to the energy density of the Universe and the cosmological acceleration ensues.
For the coupled neutrino-scalar field fluid the squared sound speed may become negative - a signal of instability \cite{Bea08}.
The natural interpretation of this instability is that the Universe becomes inhomogeneous due to nonlinear fluctuations of neutrinos, which eventually collapse into lumps \cite{Mot08}.

The question here is if the fraction corresponding to the neutrino energy contribution is due to an isotropic and homogenous NR ($p \sim 0$) distribution of particles, i. e. $M_{\0}$, or if it is due to a gas of weakly interacting particles ($p > 0$) which lump in an expanding Universe, that is $M$.
The rates given by $M_{\0}/M$ for different scenarios of mass varying mechanism leads to a reasonable estimative for the corrections on the absolute values of the cosmological neutrino masses.

So far, the known cosmological neutrino mass predictions are performed for an isotropic and homogenous NR distribution of particles \cite{Kla04,Ich06,Han05,Pas07}, which in the absence of internal interactions and gravitational forces yield a total energy (per equivalent lump) equal to $M_{\0}$.
It would be interesting to quantify the contribution of neutrino lumps in the scenarios of cosmological neutrino masses.

For the case where the total neutrino number $N$ is assumed to be conserved (clearly an approximation), one can define the apparent (measured) value for neutrino masses as $\langle \mu \rangle = M/N$ and the realistic (expected) corresponding value as $\langle m \rangle = M_{\0}/N$ so that
\begin{equation}
\frac{\langle m \rangle}{\langle \mu \rangle} = \frac{M_{\0}}{M}.
\label{star20}
\end{equation}
Astrophysical objects composed by neutrinos which similar features have also been studied in Refs. \cite{Ste96,Bil99,Tet08}.

Let us then consider that the energy density of the Universe involves weakly interacting neutrinos described by the previously discussed static, spherically symmetric solution of the Einstein equation with constant energy density.
The neutrino mass depends on the value of a slowly varying scalar field, $m\bb{\phi\bb{r}}\sim m\bb{r}$, which is responsible for the attractive force responsible for the lumping.
Independently of the analytical expression of the mass varying mechanism, we assume that the neutrino mass is equal to $m_{\0}$ at $R$, that is on the surface of the spherical lump where $p = 0$.

\begin{figure}
\vspace{-1.0 cm}
\centerline{\psfig{file=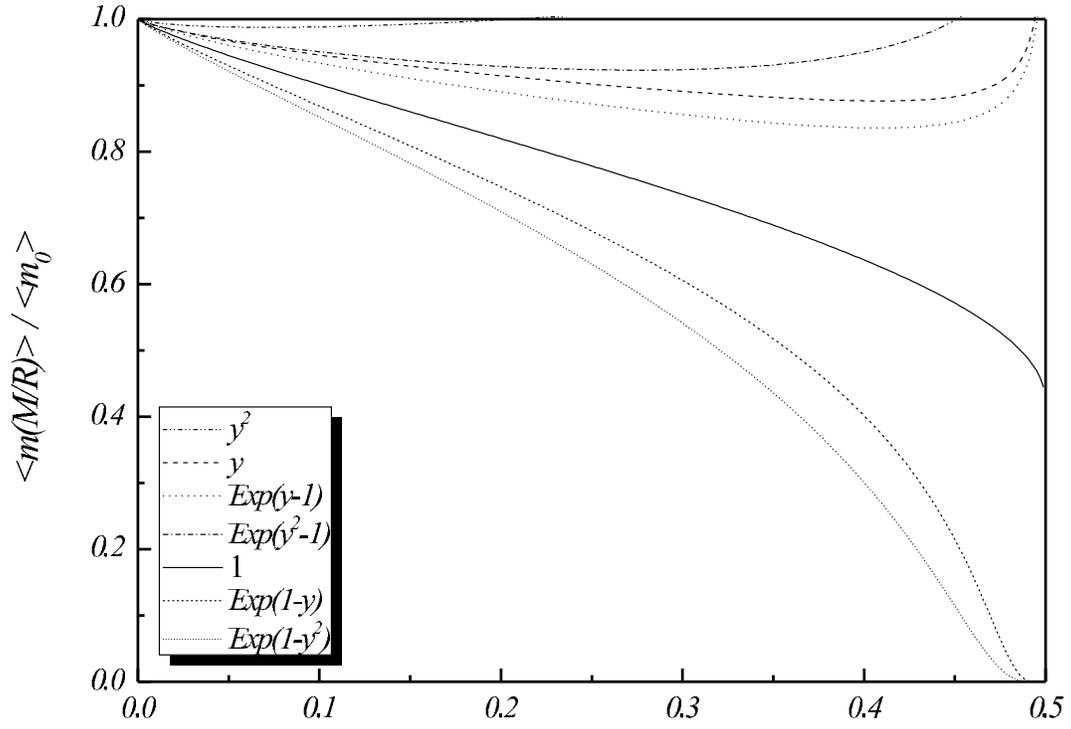, width=14cm}}
\vspace{-1.0 cm}
\caption{Modified predictions for the absolute value of neutrino masses due to the lumping of neutrinos.
For all stable scenarios the absolute neutrino mass predictions are shifted towards smaller values.}
\label{Fstar07}
\end{figure}

In the Fig.~\ref{Fstar07} we represent the rate $\langle m \rangle/\langle \mu \rangle$ for different mass varying scenarios aiming to obtain realistic values for the neutrino masses.
We observe a clear correspondence with the stability curves of the Fig.~\ref{Fstar04}.

Naturally, the smaller is the spherical radius, the more isotropic and homogeneous is the cosmological neutrino distribution.
For lumps in which the neutrino mass decreases from the boundary to the center, we have a lower limit for the rate $\langle m \rangle/\langle \mu \rangle$, which coincides with the upper limit for the radius of stable equilibrium configurations.
For the other cases, that is, for static or increasing neutrino masses from the surface to the interior, we get stable equilibrium configurations for which the rate $\langle m \rangle/\langle \mu \rangle$ decreases till vanishing at $R$.
From the analysis of equilibrium and stability, we notice that an unlimited quantity of mass energy $M$ can be squeezed into a lump when the particle mass increases inwards towards the center of the star, which is a preliminary condition to form a black-hole late configuration.
A crude interpretation of this is that the apparent neutrino mass parameter, $\langle m \rangle$, assumes smaller values in the Schwarzchild exterior solution.
Since the analytical dependence of the scalar field on $r$ is suppressed, the value of $\langle m \rangle$ can be interpreted as the cosmological neutrino mass.
Although somewhat paradoxical, this is a natural consequence of the mass varying mechanism.
Actually, for these lumps, most of the mass is concentrated at the center.
Assuming that predictions for the neutrino energy density and for the corresponding particle number density are unaltered by the lumping scenarios, the neutrino mass value vanishes at the boundary ($r=R$) of this objects if the ratio $M/R$ reaches the Buchdahl's limit.
The neutrino masses in this cases are expected to vanish.

On general terms, all stable scenarios of compact lumps lead to a shift towards smaller values for the neutrino masses.
Comparing two possible scenarios for cosmological neutrinos subject to a coupling to the background scalar field, one  with an isotropic and homogenous energy distribution, vis a vis another whose perturbations result in stable neutrino lumps, the latter correspond to lower masses at present, that is, $\lesssim 0.07 \, eV$.

\section{Conclusions}
In this work we have shown that the presence of an interaction between nonbaryonic matter, either neutrinos or dark matter, to a cosmological background scalar field associated to dark energy, has interesting implications, besides the changes on the evolution of energy density components of the Universe.

In our approach, we have assumed that any kind of dynamical mass behaves like an additional extensive thermodynamic degree of freedom, depending on the background scalar field.
Under adiabatic conditions, the thermodynamic pressure and its explicit dependence on the curved space radial coordinate could be expressed as $\rho\bb{r}$, and to be computed from the TOV equations for the hydrostatic equilibrium \cite{Tol39,Vol39}.
This justifies the assumption of an explicit analytical dependence of the variable mass, $m\bb{r}$.
Consequently, a connection among the energy density the particle variable mass and the particle density is established.
Furthermore, the analysis provides the necessary conditions to examine the equilibrium and stability of static, spherically symmetric compact structures

For cosmological scenarios which admit some kind of mass varying mechanism, the instabilities associated with an imaginary speed of sound in the NR regime do play a role in the structure formation.
Effectively, the scalar field mediates an attractive force between the nonbaryonic particles and may lead to the formation of matter lumps.
Actually, this would induce the combined fluid to form compact structures which would behave like cold dark matter.
The detection of these compact objects could be made through their gravitational potential, or through their effect on baryons.
Therefore, the existence of nonbaryonic matter lumps coupled with baryonic structures like stars or cold matter, should not be discarded \cite{Hol09}.

We have also established a direct connection between the size of neutrino lumps and the neutrino masses.

One interesting property of our solutions is that small modifications on the potential of the scalar field does not affect significantly the analysis of equilibrium and stability conditions.

Our results reinforce the argument that the Higgs \cite{Ber09} and the neutrino \cite{Ber09C} sectors are possibly the only ones where one can couple a new SM singlet without upsetting the known phenomenology.
We believe that our proposal is a further concrete step in this respect.

\begin{acknowledgments}
A. E. B. would like to thank the financial support from the Brazilian Agencies, FAPESP (grant 08/50671-0) and CNPq (grant 300627/2007-6).
\end{acknowledgments}

\pagebreak
\newpage

\end{document}